\documentclass[%
 reprint,
 amsmath,amssymb,
 aps,
]{revtex4-2}

\usepackage{graphicx}
\usepackage{dcolumn}
\usepackage{bm}
\usepackage{lmodern}
\usepackage{amsfonts}
\usepackage{subfig}
\usepackage{amsmath}
\usepackage{amssymb}
\usepackage{amsthm}
\usepackage{tensor}
\usepackage{physics}
\usepackage{stmaryrd}
\usepackage{color,soul}
\usepackage{hyperref}
\usepackage[T1]{fontenc}
\usepackage{graphicx}
\usepackage{caption}
\theoremstyle{remark}
\newtheorem{remark}{Remark}

\renewcommand{\d}{\mathrm{d}}

\newcommand{\romannumup}[1]{\uppercase\expandafter{\romannumeral#1}}
\newcommand{\romannumlow}[1]{\expandafter{\romannumeral#1}}

\begin{document}


\title{\Large\textbf{No-Go Theorem for Singularity Resolution}}

\author{Zhen-Xiao Zhang$^{1}$}
\email{zx.zhang@mail.nankai.edu.cn}
\author{Chen Lan$^{2}$}
\email{stlanchen@126.com}
\author{Yan-Gang Miao$^{1}$}
\email{Corresponding author:miaoyg@nankai.edu.cn}
	\affiliation{$^1$School of Physics, Nankai University, 94 Weijin Road, Tianjin 300071, China}
	\affiliation{$^2$Department of Physics, Yantai University, 30 Qingquan Road, Yantai 264005, China}


\begin{abstract}

We prove a No-Go theorem for singularity resolution in homogeneous, spatially flat gravitational collapse: within this sector, quantum corrections introduced solely as non-vanishing effective matter sources are insufficient to halt singularities in any vacuum-normalized analytic gravitational theory, including general relativity and other theories with analytic gravitational actions. This theorem rules out singularity resolution via effective energy density in a broad class of quantum gravity approaches, including asymptotic safety and noncommutative geometry theories, where the effective energy densities yield finite-time singularities or geodesic incompleteness. The singularity resolution strictly requires non-analytic modifications of the gravitational response at $\mathbb{Q}=0$, or a vanishing effective energy density at high densities (as realized in loop quantum gravity's Planck stars). The theorem is proved via an intrinsic $f(\mathbb{Q})$ gravity framework, extended universally to general relativity, $f(\mathbb{R})$, and $f(\mathbb{T})$ theories through the geometrical trinity at the level of the corresponding homogeneous collapse response structure—with regularity criteria and junction conditions grounded in non-metricity, free of standard GR tools.

\end{abstract}

\maketitle


\section{Introduction}

The fate of matter under gravitational collapse is one of the deepest open questions in theoretical physics. General Relativity (GR) predicts that the collapse inevitably produces spacetime singularities~\cite{Hawking:1970zqf}, the regions where the predictability of GR breaks down and GR itself fails, together with GR's classical equivalent descriptions (TEGR and STEGR, see below). The leading resolution strategy, shared by virtually all quantum gravity (QG) approaches, is to introduce quantum corrections as an effective energy density $\epsilon_\text{eff}$, an equivalent matter source that suppresses collapse at high densities~\cite{Ashtekar:2006rx}. This underlies the asymptotic safety theory~\cite{Bonanno:2000ep}, the loop quantum gravity (LQG)~\cite{Gambini:2013ooa,PhysRevD.99.124012}, and the noncommutative geometry theory~\cite{Connes:1996gi,Nicolini:2005vd,Chamseddine:2014nxa}, as well as a broad class of regular black hole models~\cite{Lan:2023cvz}. Despite their diversity, these theories and models share a common structure: they modify the matter sector while leaving the analytic structure of the gravitational action intact. By matching boundary conditions on the collapsing matter's surface, one enables matching to a static exterior geometry, often invoking generalizations of Birkhoff's theorem~\cite{Bonanno:2023rzk,Panassiti:2025diw,Zhang:2025cov}, in these theories and models. However, such a treatment relies on phenomenological and model-specific analyses, requiring case-by-case verification of regularity.  It still remains an open question whether the shared strategy of modifying the matter sector within an analytic gravitational theory can ever suffice for the genuine singularity resolution.

We address this issue within a unified framework provided by the ``geometrical trinity'' of gravity. In the geometrical trinity, the gravitational interaction can be described equivalently by curvature, torsion, or non-metricity. Replacing the curvature description of GR by a curvature-free but torsionful geometry yields TEGR, while using a curvature-free and torsion-free geometry with non-metricity~\cite{BeltranJimenez:2017tkd,Heisenberg:2018vsk} ($Q_{\alpha\mu\nu}=\nabla_\alpha g_{\mu\nu}$, where $\nabla$ denotes the general affine connection) yields STEGR. GR, TEGR, and STEGR are equivalent classical gravity descriptions whose extensions, $f(\mathbb{R})$, $f(\mathbb{T})$, and $f(\mathbb{Q})$ gravity~\cite{Heisenberg:2023lru,Zhang:2025dzt}, offer complementary mathematical perspectives on collapse. We exploit $f(\mathbb{Q})$ gravity as a particularly clean arena: its flat-spacetime formulation eliminates curvature entirely, enabling a fully intrinsic collapse framework free of GR-derived tools. The No-Go theorem we shall prove, however, is expected to apply to the corresponding analytic FLRW sectors of the geometrical trinity whenever the dynamics can be reduced to a vacuum-normalized response equation of the form used below via the geometrical trinity.

However, as extensions of GR, $f(\mathbb{T})$ and $f(\mathbb{Q})$ lack many GR-based properties, especially in the nascent $f(\mathbb{Q})$ gravity. The previous studies often borrow GR concepts (e.g., Riemann and extrinsic curvatures via the Levi-Civita connection) as auxiliaries, despite $f(\mathbb{Q})$'s distinct non-metric-compatible connection. This introduces rigor issues, as it is conceptually inappropriate to diagnose regularity by curvature invariants of the independent flat connection.

The implications of our result are broad and model-independent. For asymptotic-safety-inspired effective-source models formulated in the spatially flat Oppenheimer-Snyder sector with minimal matter-geometry coupling, our theorem implies that the collapse cannot be rendered genuinely regular without either a vanishing effective density at the endpoint or a modification of the analytic gravitational response~\cite{Bonanno:2000ep,Bonanno:2023rzk}. For the noncommutative geometry theory, the same conclusion applies. For the regular black hole model-building, our result provides a no-go criterion that bypasses the case-by-case analysis: any regular black hole, if sourced by an effective QG density within an analytic theory, must harbor either a scalar singularity or geodesic incompleteness. The theorem thus transforms singularity resolution from a model-dependent question into a matter of algebraic structure.

In this paper, leveraging $f(\mathbb{Q})$'s advantages for the collapse analysis, we establish a purely intrinsic $f(\mathbb{Q})$-based framework, including regularity criteria and junction conditions grounded in non-metricity. The regularity criteria and junction conditions recover GR results when $f(\mathbb{Q}) \to \mathbb{Q}$. Using this property, we discuss QG contributions via $\epsilon_\text{eff}$ to the dust collapse, encompassing most QG theories. We prove the No-Go theorem: $\epsilon_\text{eff}$ modifications alone fail to resolve singularities in analytic theories with analytic gravitational actions (including GR) because such theories lead to finite-time singularities or geodesic incompleteness in an asymptotic collapse. The resolution requires that the action changes its analytic form to non-analytic one at $\mathbb{Q}=0$, or that its solution has a vanishing effective energy density as in LQG. We illustrate with a quadratic $f(\mathbb{Q})$ model under the asymptotic safety. Via the geometrical trinity, our results extend to GR, $f(\mathbb{R})$, and $f(\mathbb{T})$. This suggests many QG approaches, like the asymptotic safety and noncommutative geometry, cannot eliminate singularities without fundamental ultraviolet restructuring.

\section{Formalism}

The symmetric teleparallel gravity and its $f(\mathbb{Q})$ extensions describe spacetime as a manifold with vanishing curvature $R^\rho_{\sigma\mu\nu}$ and torsion $T^\lambda_{\mu\nu}$. Instead, a general connection yields non-metricity $Q_{\alpha\mu\nu} = \nabla_\alpha g_{\mu\nu}$. Varying the action
\begin{equation}
    S = \int \d^4x \sqrt{-g} \left[ \frac{1}{2\kappa} f(\mathbb{Q}) + \mathcal{L}_m \right],
\end{equation}
gives the field equations
\begin{equation}
\begin{split}
    \frac{2}{\sqrt{-g}} \nabla_\alpha & \left[\sqrt{-g} f'(\mathbb{Q}) \tensor{P}{^\alpha_\mu_\nu} \right] \\
    + & f'(\mathbb{Q}) q_{\mu\nu} - \frac{1}{2} f(\mathbb{Q}) g_{\mu\nu} = \kappa \mathcal{T}_{\mu\nu},
    \label{eq:f(Q)_metric_field_eq}
\end{split}
\end{equation}
where $\tensor{P}{^\alpha_\mu_\nu}$, $q_{\mu\nu}$, and $\mathbb{Q}$ are defined in Sec.~\ref{sec:sm_def} in the Supplementary Material (SM).
In the limit $f(\mathbb{Q})=\mathbb{Q}$, Eq.~(\ref{eq:f(Q)_metric_field_eq}) reduces to Einstein's equations.

We adopt the coincident gauge, $\Gamma^\mu_{\nu\rho}=0$, where $Q_{\alpha\mu\nu} = \partial_\alpha g_{\mu\nu}$, with which we can simplify calculations by eliminating connection terms, allowing our direct focus on metric variations in flat coordinates.

\subsection{Intrinsic regularity criteria}

It is a non-trivial challenge to assess black hole regularity in $f(\mathbb{Q})$ gravity because the $f(\mathbb{Q})$'s non-Riemannian structure with vanishing curvature and torsion renders the direct application of GR criteria ineffective. In GR, the regularity requires finite curvature invariants (e.g., Kretschmann scalar $K=R_{\mu\nu\rho\sigma}R^{\mu\nu\rho\sigma}$) and geodesic completeness. However, in $f(\mathbb{Q})$, the Riemann tensor is identically zero, making such invariants trivially vanishing and unsuitable for detecting pathologies like divergences in tidal forces or metric deformations. The physical motivation is transparent: these invariants measure tidal distortion rates and metric deformation, playing the role of curvature invariants in GR. As we prove in the SM, the geometric finiteness of $\mathbb{Q}$ and $\nabla Q$ invariants implies the finiteness of the Kretschmann scalar, recovering the GR criterion in the appropriate limit.

To address the regularity, we propose an intrinsic framework grounded in $f(\mathbb{Q})$'s fundamental variable: the non-metricity tensor $Q_{\alpha\mu\nu}$ and its covariant derivatives. A black hole solution is intrinsically regular if the following two conditions hold:
\begin{enumerate}
    \item \textbf{Geometric Finiteness}: All independent scalar invariants constructed from $Q_{\alpha\mu\nu}$ and $\nabla_\beta Q_{\alpha\mu\nu}$ remain finite everywhere.
    The key examples include $\mathbb{Q}$, $Q_\alpha Q^\alpha$, $\overline{Q}_\alpha \overline{Q}^\alpha$, and higher-order contractions like $\nabla_\beta Q_{\alpha\mu\nu} \nabla^\beta Q^{\alpha\mu\nu}$.
    They capture metric deformation rates, analogous to curvature invariants in GR,
    and diverge at singularities. 
    \item \textbf{Geodesic Completeness}: All timelike/null geodesics defined by the Levi-Civita connection 
    $\tensor{\mathring{\Gamma}}{^\mu_\nu_\rho}$
    (derived solely from the metric $g_{\mu\nu}$) are complete,
    where the affine parameter $\lambda$ is extendable to infinity:
    $\d^2 x^\mu/\d \lambda^2 + \tensor{\mathring{\Gamma}}{^\mu_\nu_\rho} (\d x^\nu/ \d \lambda)(\d x^\rho / \d \lambda)=0$. 
    This choice is mandated by the principle of minimal coupling, which ensures that matter fields couple exclusively to the metric, rendering the affine connection physically irrelevant for particle trajectories. Furthermore, since the Levi-Civita connection depends solely on the metric, this criterion is strictly invariant under the inertial gauge transformations of the theory (Further discussed in the SM). 
\end{enumerate}
The above approach is rigorous and self-consistent: it recovers the convergence of GR invariants (e.g., the Kretschmann scalar, see the proof in the SM), avoids inappropriate curvature computations in spacetime where the curvature vanishes, and aligns with $f(\mathbb{Q})$'s metric-affine foundations. It enables unambiguous regularity checks for collapse solutions, highlighting $f(\mathbb{Q})$'s utility over GR-borrowed tools.

\subsection{Intrinsic junction conditions}
Analogous to Israel's conditions in GR, for hypersurface $\Sigma$ separating $\mathcal{M}_+$ and $\mathcal{M}_-$, we propose
\begin{enumerate}
    \item Induced metric continuous: $\llbracket h_{\mu\nu}\rrbracket=0$;
    \item Normal component of a gravitational conjugate momentum continuous:
    $\llbracket n_\alpha \tensor{\Pi}{^\alpha_\mu_\nu}\rrbracket=\kappa S_{\mu\nu}/2$,
\end{enumerate}
where the gravitational conjugate momentum is defined as $\tensor{\Pi}{^\alpha_\mu_\nu} = \sqrt{-g} f'(\mathbb{Q}) \tensor{P}{^\alpha_\mu_\nu}$, $n_\alpha$ is the normal covector, and $S_{\mu\nu}$ the surface stress-energy. Here $\llbracket\cdot\rrbracket$ represents the discontinuity across the hypersurface. The second condition ensures the conservation of momenta  conjugate to the metric, where the gravity Lagrangian density is given by $\mathcal{L}_\text{grav}=\sqrt{-g}f(\mathbb{Q})/(2\kappa)$ and $\Pi^\alpha{}_{\mu\nu}=\sqrt{-g}f'(\mathbb Q)P^\alpha{}_{\mu\nu}$ is a rescaled momentum density proportional to the canonical momentum $\partial \mathcal L_{\rm grav}/\partial Q_{\alpha\mu\nu}$. It recovers Israel's second condition when $f(\mathbb{Q})=\mathbb{Q}$ (See the proof in the SM).

\section{Dynamical obstruction and No-Go theorem} \label{sec:main_part}

The strategy of our proof is as follows. The most general scenario is that of homogeneous dust collapse, where QG contributes an effective matter source. It is modeled by a contracting Friedmann-Lemaître-Robertson-Walker (FLRW) interior matched to a static exterior. In this setting, the entire collapse dynamics is encoded in a single function: the scale factor $a(t)$. The modified Friedmann equation in $f(\mathbb{Q})$ gravity then takes the form of a master equation $\mathcal{J}(\mathbb{Q}) = \epsilon_{\rm eff}(a)$, which separates the gravitational response $\mathcal{J}$ from the quantum-corrected matter source $\epsilon_{\rm eff}$. Singularity resolution requires either (i) a bounce, meaning $\dot{a}=0$ at some finite $a_{\rm min}>0$, or (ii) asymptotic non-collapse, meaning $a(t)\to{\rm const.}>0$. We show that both possibilities are obstructed for any analytic $f(\mathbb{Q})$ including GR: case (i) is forbidden by the algebraic identity $\mathcal{J}(0)=0$, and case (ii) is possible in the analytic sector only if the effective density vanishes at the limiting radius; it is therefore excluded for a strictly positive effective density.

To rigorously analyze the singularity resolution, we adopt the Oppenheimer-Snyder approach, which models the interior of a collapsing homogeneous dust cloud in terms of a contracting FLRW metric. Such a way naturally dictates that the collapse dynamics is governed by the Friedmann equations. We separate~\cite{Heisenberg:2023lru} gravity from matter by rewriting the Friedmann equation in $f(\mathbb{Q})$ gravity, which yields the master equation for the collapse dynamics with scale factor $a(t)$
\begin{equation}
    \mathcal{J}(\mathbb{Q})=\epsilon_\text{eff}(a),
    \label{eq:master_eq}
\end{equation}
where the dynamical response is 
\begin{equation}
    \mathcal{J}(\mathbb{Q})\equiv \mathbb{Q}f'(\mathbb{Q})-\frac{1}{2}f(\mathbb{Q}), 
\end{equation}
and $\epsilon_\text{eff}(a)$ is the effective energy density from collapsing dust, defined as $\epsilon_\text{eff}(a)=\kappa\rho_\text{eff}(a)$. In the coincident gauge, $\mathbb{Q} = 6 H^2\ge 0$, with Hubble parameter $H=\dot{a} / a$.

~\\

\begin{remark}
    \textit{In this work, an analytic $f(\mathbb{Q})$ means that $f(\mathbb{Q})$ admits a convergent Taylor expansion in an open interval containing $\mathbb{Q}=0$, which has no pole or singularity along the collapse trajectory. Thus logarithmic terms, fractional powers and rational functions with poles at $\mathbb{Q}=0$ are not analytic in the sense used in this work.}
\end{remark}

~\\

\begin{remark}
    \textit{The no-go statement below is restricted to the homogeneous, isotropic and spatially flat Oppenheimer--Snyder sector, where the collapse dynamics is described by a single scale factor $a(t)$. We also assume minimal matter--geometry coupling and an effective-source description in which the quantum correction enters through a vacuum-subtracted effective density $\bar\epsilon_{\rm eff}(a)$. Positively curved $K=1$ collapse, rotating or shearing configurations, and models with explicit non-minimal matter--geometry couplings are therefore outside the proof given here.}
\end{remark}

~\\

We analyze collapse by focusing on singularity-free scenarios in finite time (requiring detailed asymptotic checks). Two possibilities emerge:
\begin{enumerate}
    \item Asymptotic collapse to zero ($a(t)\to 0$ as $t\to\infty$);
    \item Bounce or stabilization at finite radius ($a(t)\ge a_{\rm min} > 0$).
\end{enumerate}

\subsection{Scenario 1: Asymptotic collapse without bounce ($a(t)\to 0$)}

We examine $a(t)\to0$ by distinguishing asymptotic $\epsilon_\text{eff}$ behaviors:
\begin{itemize}
    \item Unbounded source: $\lim_{a\to 0}\epsilon_\text{eff}(a) = \infty$. 
    Eq.~\eqref{eq:master_eq} leads to $\mathcal{J}(\mathbb{Q})\to \infty$; for an analytic $f(\mathbb{Q})$,
    this implies $\mathbb{Q} \to \infty$, a scalar singularity.
    \item Saturated source: $\lim_{a\to 0}\epsilon_\text{eff}(a) = \epsilon_\text{max} < \infty$.
    $\mathbb{Q} \to \mathbb{Q}_{\rm const.}$, yielding de Sitter-like contraction: $a(t)\sim \exp(-t\sqrt{\mathbb{Q}_{\rm const.}/6})$ as $t\to\infty$.
    The null geodesic affine parameter, $\lambda(t) \sim \int a(t) \d t \sim \exp(-t\sqrt{\mathbb{Q}_{\rm const.}/6})$, converges to a finite value,
    implying incompleteness for all future-directed radial geodesics (See the proof in the SM).
\end{itemize}
Thus, the asymptotic collapse without bounce always yields singularity, either to a scalar singularity or to geodesic incompleteness.

\subsection{Scenario 2: Bounce or Asymptotic Saturation}

A bounce requires $\dot{a}=0$ at finite $a\neq 0$, so $\mathbb{Q}=0$ and $\mathcal{J}(0)=\epsilon_\text{eff}(a)$. Unless the QG model allows $\epsilon_\text{eff}(a)=0$, we have $\epsilon_\text{eff}(a)>0$. For a vacuum-normalized analytic $f(\mathbb Q)$, its Taylor expansion near $\mathbb Q=0$ takes the form
\begin{equation}
    f(\mathbb Q)=\sum_{n=0}^{\infty}c_n\mathbb{Q}^n .
\end{equation}

Then
\begin{equation}
    \mathcal{J}(0)=\sum_i\left(i-\frac{1}{2}\right)c_i\mathbb{Q}^i\bigg|_{\mathbb{Q}=0}=-\frac{c_0}{2}.
\end{equation}
With $\epsilon_\text{eff}>0$, no bounce solution exists. The reason is that each term, $c_i \mathbb{Q}^i$, contributes $(i - 1/2)c_i \mathbb{Q}^i$ to $\mathcal{J}(\mathbb{Q})$; at $\mathbb{Q}=0$, only the $i=0$ constant term survives, so that $\mathcal{J}(0)=-c_0/2$, which vanishes in the case of asymptotically flat spacetime due to $c_0=0$. While for non-asymptotically flat spacetime, with a finite constant term $f_0=f(0)$, the response at the bounce is $\mathcal J(0)=-f_0/2$. Such a term is equivalent to a vacuum-energy shift. After defining $\bar f(\mathbb Q)=f(\mathbb Q)-f_0$ and $\bar\epsilon_{\rm eff}(a)=\epsilon_{\rm eff}(a)+f_0/2$, the master equation becomes $\bar{\mathcal J}(\mathbb Q)=\bar\epsilon_{\rm eff}(a)$, with $\bar{\mathcal J}(0)=0$. Hence a regular finite-radius bounce in an analytic theory requires $\bar\epsilon_{\rm eff}(a_b)=0$. If the vacuum-subtracted effective density remains positive, the bounce is still forbidden. In this sense the assumption $f(0)=0$ is not essential; it merely corresponds to the asymptotically flat or vacuum-subtracted normalization.

\textbf{No-Go Theorem:} \textit{In the gravitational theory with analytic gravitational action term, under homogeneous, spatially flat sectors,} singularities produced by gravitational collapse cannot be resolved solely by quantum-gravity corrections encoded as a strictly positive effective energy density. A successful resolution requires either a vanishing effective density at the would-be regular endpoint or a non-analytic gravitational action term.

A necessary bounce condition is $\lim_{\mathbb{Q}\to 0} \mathcal{J}(\mathbb{Q})>0$. If $\mathcal{J}(0) = \lim_{t\to\infty} \epsilon_\text{eff}(a(t))$, the bounce occurs at infinity: e.g., $a(t)$ approaches constant $a_\infty>0$ as $t\to \infty$, a `frozen' regular state interpretable as an asymptotic saturation, or an infinite-time bounce. The bounce regularity is verified via our criteria (See the details in the SM). 

We summarize for QG with effective matter as follows:
\begin{enumerate}
    \item Non-vanishing $\epsilon_\text{eff}$ requires a gravitational action term $f(\mathbb{Q})$ which is non-analytic at $\mathbb{Q}=0$, that satisfies $\lim_{\mathbb{Q}\to 0}\mathcal{J}(\mathbb{Q})>0$, for singularity resolution;
    \item Analytic frameworks succeed only if $\epsilon_\text{eff} \to 0$ (e.g., $\epsilon_\text{eff}=\epsilon(1-\epsilon/\epsilon_0)$ in LQG), yielding Planck stars~\cite{2014IJMPD..2342026R}.
\end{enumerate}

While applying to most analytic QG, a non-analytic $f(\mathbb{Q})$ at $\mathbb{Q}=0$ may enable bounces if $\lim_{\mathbb{Q}\to 0} \mathcal{J}(\mathbb{Q})>0$ holds, though the ultraviolet viability needs scrutiny.

\begin{figure}
    \centering
    \includegraphics[width=\linewidth]{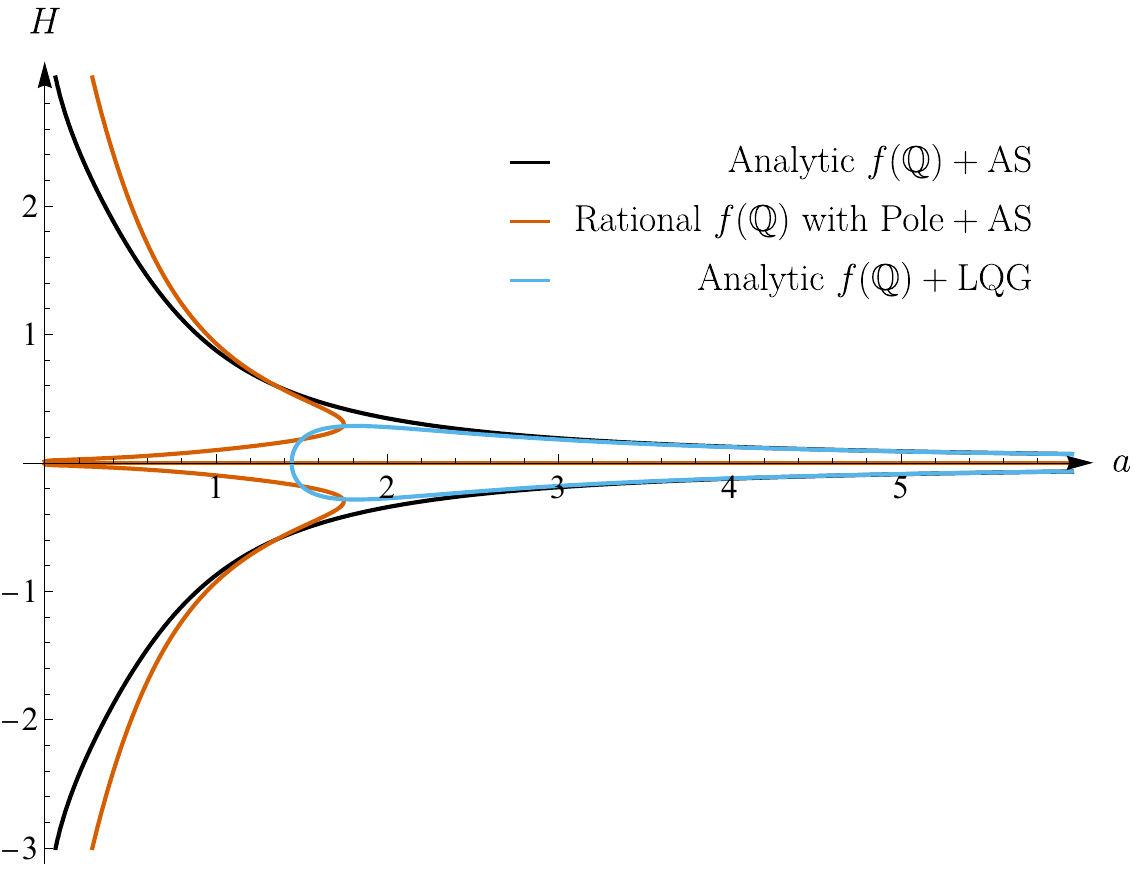}
\captionsetup{justification=raggedright}
\caption{$a-H$ phase diagram for collapse progress. The upper ($H>0$) and lower ($H<0$) branches 
represent the expansion and collapse processes, respectively. 
(a) the combination of an analytic $f(\mathbb{Q})$ ($f(\mathbb{Q})=\mathbb{Q}+\alpha \mathbb{Q}^2$, with $\alpha=0.01$) with asymptotic safety (black); 
(b) the combination of a rational $f(\mathbb{Q})$ correction with a pole at $\mathbb{Q}=0$} ($f(\mathbb{Q})=\mathbb{Q}-\lambda/\mathbb{Q}$, with $\lambda=0.1$) with asymptotic safety (red); 
(c) the combination of an analytic $f(\mathbb{Q})$ ($f(\mathbb{Q})=\mathbb{Q}+\alpha \mathbb{Q}^2$, with $\alpha=0.01$) with LQG (blue). 
In the first case, the system is strictly divided into upper and lower branches. In the latter two cases, the collapsing and expanding branches are connected.
\label{fig:phase_graph}
\end{figure}

As an example of typical scenarios, we present in Fig.~\ref{fig:phase_graph} the $a-H$ phase diagrams for three representative cases. The conditions under which a bounce may occur show that the upper ($H>0$) and lower ($H<0$) branches of the figure are connected, where the two branches represent expansion and collapse processes, respectively. These cases include the combination of an analytic $f(\mathbb{Q})$ ($f(\mathbb{Q})=\mathbb{Q}+\alpha \mathbb{Q}^2$) with asymptotic safety theory, the combination of a rational $f(\mathbb{Q})$ correction with a pole at $\mathbb{Q}=0$($f(\mathbb{Q})=\mathbb{Q}-\lambda/\mathbb{Q}$) with asymptotic safety theory, and the combination of an analytic $f(\mathbb{Q})$ ($f(\mathbb{Q})=\mathbb{Q}+\alpha \mathbb{Q}^2$) with LQG. Our theoretical predictions suggest that the first case should not admit bounce solutions, while the latter two cases may exhibit bounce solutions. Consistent with theoretical expectations, the phase diagrams show that the system in the first case remains confined to the $H<0$ region for collapsing initial conditions, making the collapse unstoppable. In the latter two cases, however, the system can evolve to the $H=0$ state and cross the horizontal axis, enabling a bounce to occur.

\section{Example: Quadratic Analytic Gravity with Asymptotic-Safety Matter}

To illustrate the general obstruction derived above, we examine the dust collapse in the spatially flat Oppenheimer-Snyder model~\cite{Oppenheimer:1939ue}:
\begin{equation}
    \epsilon=\frac{3m_0}{a(t)^3},\qquad p=0,
\end{equation}
with an Asymptotic Safety correction~\cite{Bonanno:2023rzk}:
\begin{equation}
    \epsilon_\text{eff}=\frac{1}{\xi}\ln\left(\frac{3m_0\xi}{a^3}+1\right).
\end{equation}
This unbounded source ($\epsilon_\text{eff}>0$, $\lim_{a\to 0}\epsilon_\text{eff}=\infty$) should yield a singularity as predicted in Sec.~\ref{sec:main_part}. It is notable that other AS-inspired implementations, where additional physical modifications are regarded as non-minimal coupling assumptions, lie outside the present example. 

We adopt a quadratic action
\begin{equation}
    f(\mathbb{Q})=\mathbb{Q}+\alpha\mathbb{Q}^2,
\end{equation} 
which recovers GR as $\alpha\to 0$. Here, due to $\mathcal{J}(0)=0$, together with $\epsilon_\text{eff}>0$, the bounces are forbidden, aligning with the theorem.

The modified Friedmann equation gives
\begin{equation}
    \dot{a}(t)=-\frac{a(t)}{6} \sqrt{\frac{\sqrt{\frac{24\alpha}{\xi}  \ln \left(\frac{3 m_0 \xi }{a(t)^3}+1\right)+1 }-1}{\alpha}}.
    \label{eq:quad_adott}
\end{equation}

\subsection{Failure to Eliminate Singularity}

The asymptotic behavior of $a(t)$ for large $t$ is given by (see derivation in SM)
\begin{equation}
    a(t)\sim D_0 \exp \left[-E_0 \left(C_0 t-C_1\right)^\frac{4}{3} \right],
    \label{eq:approx_a}
\end{equation}
where
\begin{equation}
    C_0=\frac{1}{6}\sqrt[4]{\frac{24}{\alpha  \xi }},\qquad D_0=\sqrt[3]{3 m_0 \xi},\qquad E_0=\sqrt[3]{\frac{3^{5}}{2^{8}}}.
\end{equation}
The parameter $C_1$ is the integration constant determined by the initial scale of the dust cloud. This is precisely the scenario of asymptotic collapse towards zero that we discussed above. 

Evaluating the non-metricity scalar $\mathbb{Q}$ as $t\to\infty$, we find
\begin{equation}
    \lim_{t\to\infty}\mathbb{Q}=\lim_{t\to\infty}6\left[\frac{\dot{a}(t)}{a(t)}\right]^2=\lim_{a\to 0}C \left[ \ln \left(\frac{1}{a}\right) \right]^{1/2}
\end{equation}
diverges, where $C=\sqrt{2/\alpha\xi}$ is constant. This indicates that the solution lacks regularity. 

Furthermore, even apart from the scalar divergence, the spacetime remains geodesically incomplete as defined by our Intrinsic Regularity Criteria. Consider a radial null geodesic as $t\to\infty$
\begin{equation}
    \lim_{t\to\infty}\lambda(t)\leq \int_0^{\infty} \exp\left( - k\tau^\frac{4}{3} \right) \,\d\tau=\frac{\Gamma \left(\frac{7}{4}\right)}{k^{3/4}},
\end{equation}
where $k$ is a finite positive parameter. The convergence of the affine parameter confirms geodesic incompleteness. 

Consequently, both regularity criteria fail in the analytic model: a scalar singularity ($\mathbb{Q}\to\infty$) and geodesic incompleteness. This example illustrates the no-go theorem derived in Sec.~\ref{sec:main_part}: within the analytic sector, a strictly positive effective density cannot generate a regular bounce. Singularity resolution would require either a vanishing effective density at the would-be bounce or a gravitational response outside the analytic class, whose regularity must then be checked separately.

\subsection{Junction to the Exterior Geometry}

It is worth noting that, the static solution from the perspective of an external observer can be obtained through our defined junction condition (see derivation in SM):
\begin{equation}
    M(R)=\frac{R^3\left[\sqrt{\frac{24\alpha}{\xi}  \ln \left(\frac{3 m_0 \xi  r_b^3}{R^3}+1\right)+1}-1\right]}{72 \alpha},
    \label{eq:exterior_mass_func}
\end{equation}
where $R$ is the areal radius coordinate of the exterior static metric. Here $r_b$ is the initial comoving radius of the collapsing boundary. 

Although the matching procedure formally yields a static exterior metric for a collapsing dust cloud that cannot reach a singularity within a finite time, this solution masks a fundamental pathology. The singularity is merely pushed to infinite coordinate time. Given the divergence of internal geometric scalars and the incompleteness of geodesics, the resulting spacetime fails to meet the criteria for a regular black hole. 

This finding clarifies an apparent tension with Ref.~\cite{Bonanno:2023rzk}, which reports a static regular exterior metric for an asymptotic-safety-corrected collapse. Our analysis shows that the interior spacetime remains singular while the exterior metric is indeed well-behaved and the junction condition is formally satisfied: the non-metricity scalar diverges and null geodesics are incomplete. The regular black hole of Ref.~\cite{Bonanno:2023rzk} is therefore regular only from the perspective of external observers; the interior singularity is masked, not resolved.

\section{Conclusion}

In this paper, we establish a rigorous and intrinsic framework in $f(\mathbb{Q})$ gravity for analyzing gravitational collapse, avoiding GR-derived tools by defining regularity criteria and junction conditions based on non-metricity and momentum conservation. This self-consistent approach recovers GR limits and enables generalized studies of singularity resolution.

Applying the above-mentioned framework, we propose our No-Go theorem in the homogeneous, spatially flat sector with minimal matter-geometry coupling: In the gravitational theory with analytic gravitational action term, singularities produced by gravitational collapse cannot be resolved solely by quantum-gravity corrections encoded as a strictly positive effective energy density. A successful resolution requires either a vanishing effective density at the would-be regular endpoint or a non-analytic gravitational action term. Otherwise, the dynamical response vanishes at zero expansion ($\mathcal{J}(0)=0$), obstructing bounces and leading to finite-time singularities or asymptotic collapse with divergent invariants and geodesic incompleteness—persisting even if masked for external observers.

Within the assumptions of the theorem, the non-singular resolution requires a non-singular bounce~\cite{Bojowald:2001xe}, achievable only through (i) non-analytic action modifications at ultraviolet scales, or (ii) vanishing $\epsilon_\text{eff}$ at high densities (as in LQG's Planck stars). These constraints guide future regular black hole constructions, giving implications for evaporating black holes and potential observables like modified Hawking radiation.

Our results call to mind a known limitation in constructing regular black holes via Non-Linear Electrodynamics (NLED). While the action of regular metrics can be mathematically obtained by reverse-engineered NLED Lagrangians (the Ayon-Beato and Garcia approach~\cite{Ayon-Beato:1998hmi}), a physically realistic model like the Euler-Heisenberg theory generally fails to resolve singularities~\cite{Yajima:2000kw}. Similarly, we find here that the QG theories with an effective matter source do not automatically guarantee regularity; in fact, under an analytic gravitational action and a positive effective density, they still lead to scalar singularities or geodesic incompleteness. This reinforces the conclusion that it requires more than just a modified source to resolve the singularity problem but a fundamental change in the gravitational sector itself~\cite{Carballo-Rubio:2024dca}.

We emphasize that the present theorem is not a theorem for generic gravitational collapse. In terms of future prospects, while our current analysis establishes the fundamental dynamical constraints under spherical symmetry, the realistic astrophysical collapse inevitably involves angular momentum. It is crucial to extend the present intrinsic framework to an axially symmetric spacetime for confronting theoretical predictions with upcoming observational data, such as black hole shadows~\cite{PhysRevD.108.104004} and gravitational wave ringdowns~\cite{Cardoso:2016rao}. However, this extension presents significant theoretical challenges. In particular, it is considerably more intricate to formulate the intrinsic junction conditions for rotating collapse than that for the static case in $f(\mathbb{Q})$ gravity. The matching of the generalized momentum flux across a rotating boundary involves complex non-diagonal components of the non-metricity tensor, requiring a rigorous generalization of the matching procedure developed here. Addressing these complexities constitutes a primary direction for our future research.

Via the geometrical trinity, the same obstruction exists in GR and in the corresponding analytic response sectors of $f(T)$ and $f(Q)$ gravity. For theories whose collapse equations contain additional independent dynamical variables, the extension should be understood as conditional on reducing the dynamics to the same vacuum-normalized master-equation structure. This mandates a paradigm shift in QG phenomenology toward fundamental non-analytic frameworks or vanishing $\epsilon_\text{eff}$ models. Investigating non-analytic actions offers promising directions. More broadly, our No-Go theorem reframes the singularity problem as a question about the analytic structure of gravity rather than the choice of matter source. The widespread practice of constructing regular black holes by sourcing GR with a carefully chosen $\epsilon_{\rm eff}$, whether from QG, nonlinear electrodynamics, or phenomenological models, cannot produce genuine regularity within analytic theories. The singular core is at best displaced to infinite coordinate time, remaining present as geodesic incompleteness. The true resolution demands a departure from perturbative and analytic gravity: precisely the kind of non-perturbative ultraviolet completion, such as LQG. Our results thus supply a sharp theoretical criterion for distinguishing genuine from apparent singularity resolution, with direct relevance to black hole phenomenology in the era of gravitational-wave and Event Horizon Telescope observations.


\section*{Acknowledgements}

The authors would like to express their sincere gratitude to Dr.\ Cong Zhang (Beijing Normal University) and Antonio Panassiti (University of Catania) for their valuable comments on this work.
This work was supported in part by the National Natural Science Foundation of China under Grant No.\ 12175108. C.L. is also supported by Yantai University under Grant No.\ WL22B224.
Z.-X. Z is supported by the Pilot Scheme of Talent Training in Basic Sciences (Boling Class of Physics, Nankai University), Ministry of Education.


\bibliography{references}

\clearpage
\onecolumngrid
\begin{center}
    \Large\textbf{Supplementary Material}
\end{center}
\setcounter{section}{0}
\setcounter{equation}{0}
\setcounter{figure}{0}

\section{Definitions of Basic Quantities in $f(\mathbb{Q})$ Gravity}
\label{sec:sm_def}

In $f(\mathbb{Q})$ gravity, the basic geometric quantities are defined as follows.
The non-metricity scalar $\mathbb{Q}$ is given by
\begin{equation}
\label{eq:nonmetricity}
\mathbb{Q}=P^{\alpha\mu\nu}Q_{\alpha\mu\nu},
\end{equation}
where the non-metricity conjugate $\tensor{P}{^\alpha_\mu_\nu}$ is defined as
\begin{equation}
\label{eq:conjNonmet}
\tensor{P}{^\alpha_\mu_\nu}
= \frac{1}{4}
\left(
-\tensor{Q}{^\alpha_\mu_\nu}
+2\tensor{Q}{_(_\mu^\alpha_\nu_)}
+Q^\alpha g_{\mu\nu}
-\overline{Q}^\alpha g_{\mu\nu}
-\tensor{\delta}{^\alpha_(_\mu}Q_{\nu)}
\right).
\end{equation}
Two vector quantities can be obtained from contractions of the non-metricity tensor,
\begin{equation}
Q_\alpha=\tensor{Q}{_\alpha^\lambda_\lambda},
\qquad
\overline{Q}_\alpha=\tensor{Q}{^\lambda_\alpha_\lambda}.
\end{equation}
For our later convenience, we further introduce
\begin{equation}
\label{eq:qtensor}
q_{\mu\nu}
=
P_{(\mu|\alpha\beta}Q_{\nu)}{}^{\alpha\beta}
-
2P^{\alpha\beta}{}_{(\nu}Q_{\alpha\beta|\mu)},
\end{equation}
as well as
\begin{equation}
g=\det(g_{\mu\nu}).
\end{equation}
Unless otherwise specified, we adopt the coincident gauge, in which the affine connection vanishes,
\begin{equation}
\Gamma^\mu_{\nu\rho}=0,
\end{equation}
so that the non-metricity tensor reduces to
\begin{equation}
Q_{\alpha\mu\nu}=\partial_\alpha g_{\mu\nu}.
\end{equation}

\section{Proof that Geometric Finiteness Reduces to the Kretschmann Convergence Condition in GR}

In $f(\mathbb{Q})$ gravity there exist two connections: the Levi-Civita connection $\tensor{\mathring{\Gamma}}{^\lambda_\mu_\nu}$, determined by the metric as in GR, and the independent affine connection $\tensor{\Gamma}{^\lambda_\mu_\nu}$, which vanishes in the coincident gauge adopted here. The disformation tensor, defined as the difference between these two connections, is therefore
\begin{equation}
\tensor{L}{^\lambda_\mu_\nu}
=
-\tensor{\mathring{\Gamma}}{^\lambda_\mu_\nu}
=
\frac{1}{2}
\left(
\tensor{Q}{^\lambda_\mu_\nu}
-
\tensor{Q}{_\mu^\lambda_\nu}
-
\tensor{Q}{_\nu^\lambda_\mu}
\right).
\end{equation}
It follows that $\tensor{L}{^\lambda_\mu_\nu}$ is a linear combination of the components of $\tensor{Q}{_\lambda_\mu_\nu}$, which we denote schematically as
\begin{equation}
L\sim Q.
\end{equation}
The Riemann tensor constructed from the Levi-Civita connection,
$\tensor{\mathring{R}}{^\rho_\sigma_\mu_\nu}$, is given by
\begin{equation}
\begin{aligned}
\tensor{\mathring{R}}{^\rho_\sigma_\mu_\nu}
&=
\partial_\mu\tensor{\mathring{\Gamma}}{^\rho_\nu_\sigma}
-
\partial_\nu\tensor{\mathring{\Gamma}}{^\rho_\mu_\sigma}
+
\tensor{\mathring{\Gamma}}{^\rho_\mu_\lambda}
\tensor{\mathring{\Gamma}}{^\lambda_\nu_\sigma}
-
\tensor{\mathring{\Gamma}}{^\rho_\nu_\lambda}
\tensor{\mathring{\Gamma}}{^\lambda_\mu_\sigma} \\
&=
-\nabla_\mu\tensor{L}{^\rho_\nu_\sigma}
+
\nabla_\nu\tensor{L}{^\rho_\mu_\sigma}
+
\tensor{L}{^\rho_\mu_\lambda}
\tensor{L}{^\lambda_\nu_\sigma}
-
\tensor{L}{^\rho_\nu_\lambda}
\tensor{L}{^\lambda_\mu_\sigma}.
\end{aligned}
\end{equation}
Since $L\sim Q$, it follows that
\begin{equation}
\nabla L\sim \nabla Q,
\qquad
L\cdot L\sim Q\cdot Q.
\end{equation}
Consequently, the Riemann tensor can be schematically expressed as
\begin{equation}
\mathring{R}
\sim
\mathcal{A}(\nabla Q)
+
\mathcal{B}(Q\cdot Q),
\end{equation}
where $\mathcal{A}$ and $\mathcal{B}$ denote linear operators involving only the metric and index contractions.
The Kretschmann scalar is defined as
\begin{equation}
K
=
\mathring{R}_{\mu\nu\rho\sigma}
\mathring{R}^{\mu\nu\rho\sigma}.
\end{equation}
Substituting the schematic structure of $\mathring{R}$ yields
\begin{equation}
K
\sim
(\nabla Q + Q\cdot Q)\cdot(\nabla Q + Q\cdot Q).
\end{equation}
Therefore, the Kretschmann scalar consists of a linear combination of the following three types of contributions:

\begin{enumerate}
\item $(\nabla Q)\cdot(\nabla Q)$, representing quadratic invariants of $\nabla Q$;
\item $(Q\cdot Q)\cdot(Q\cdot Q)$, representing quartic invariants of $Q$;
\item $(\nabla Q)\cdot(Q\cdot Q)$, representing mixed contractions between $\nabla Q$ and $Q\cdot Q$.
\end{enumerate}

According to our definition of geometric finiteness, the complete scalar bases constructed from both $Q$ and $\nabla Q$ remain finite. This condition guarantees that all higher-order contractions built from these quantities are finite as well. In particular, the first two classes of terms are manifestly finite. Moreover, by the Cauchy–Schwarz inequality, if the norms of $\nabla Q$ and $Q\cdot Q$ are finite, then their cross contraction $(\nabla Q)\cdot(Q\cdot Q)$ must also remain finite. Consequently, every contribution entering the Kretschmann scalar is finite, and therefore the Kretschmann scalar $K$ itself is finite.

\section{On the Use of Levi-Civita Geodesics in $f(\mathbb{Q})$ Gravity}

In $f(\mathbb{Q})$ gravity two connections coexist: the independent affine connection $\tensor{\Gamma}{^\lambda_\mu_\nu}$ and the Levi-Civita connection $\tensor{\mathring{\Gamma}}{^\lambda_\mu_\nu}$ constructed from the metric. Correspondingly, two possible notions of geodesics can be defined. The first is the autoparallel associated with the affine connection,
\begin{equation}
\ddot{x}^\lambda + \tensor{\Gamma}{^\lambda_\mu_\nu}\dot{x}^\mu \dot{x}^\nu = 0 ,
\end{equation}
while the second is the metric geodesic defined by the Levi-Civita connection,
\begin{equation}
\ddot{x}^\lambda + \tensor{\mathring{\Gamma}}{^\lambda_\mu_\nu}\dot{x}^\mu \dot{x}^\nu = 0 .
\end{equation}

In this work we argue that the physical matter follows the geodesics associated with the Levi-Civita connection. In other words, the motion of physical particles is completely determined by the spacetime metric. For this reason, the convergence or divergence of affine parameters along Levi-Civita geodesics is adopted in this work as the criterion for black hole regularity.

The choice of physical trajectories is not dictated solely by the geometric structure of the gravitational sector, but rather by how matter couples to gravity. In the standard model and in the dust collapse model considered here, the coupling follows the principle of minimal coupling. According to this principle, matter fields interact with the geometry only through the metric and do not couple directly to the affine connection. The matter action therefore takes the form
\begin{equation}
S_m=\int \d^4x \sqrt{-g}\,\mathcal{L}_m(g_{\mu\nu},\psi_m,\partial\psi_m).
\end{equation}
Consequently, the affine connection does not appear explicitly when we derive the equations of motion for particles from the matter action. For a classical point particle the action is
\begin{equation}
S_p=-m\int\d\tau\sqrt{-g_{\mu\nu}\dot{x}^\mu\dot{x}^\nu}.
\end{equation}
Since this action has the same form as in GR, varying the worldline $x^\mu(\tau)$ leads directly to the metric geodesic equation involving the Levi-Civita connection.

For continuous matter fields, the variation with respect to the metric defines the energy-momentum tensor $T^{\mu\nu}$. The Levi-Civita derivative $\mathring{\nabla}_\mu$ is the unique metric-compatible covariant derivative, which satisfies
\begin{equation}
\mathring{\nabla}_\alpha g_{\mu\nu}=0 .
\end{equation}
Only with this derivative operator can the standard conservation law
\begin{equation}
\mathring{\nabla}_\mu T^{\mu\nu}=0
\end{equation}
be obtained from the diffeomorphism invariance of the matter action. This conservation law determines the motion of matter. For example, for the dust with
\begin{equation}
T^{\mu\nu}=\rho u^\mu u^\nu ,
\end{equation}
one obtains directly
\begin{equation}
u^\mu \mathring{\nabla}_\mu u^\nu =0 ,
\end{equation}
which is precisely the Levi-Civita geodesic equation.

For completeness, we outline the derivation. Consider the variation generated by a diffeomorphism with vector field $\xi^\mu$,
\begin{equation}
\delta_\xi g_{\mu\nu}=\mathcal{L}_\xi g_{\mu\nu}.
\end{equation}
The variation of the matter action is
\begin{equation}
\delta_\xi S_m
=
\int \d^4x
\frac{\delta S_m}{\delta g_{\mu\nu}}
\delta_\xi g_{\mu\nu}
=
\frac{1}{2}
\int \d^4x\sqrt{-g}
T^{\mu\nu}
\mathcal{L}_\xi g_{\mu\nu}.
\end{equation}
The Lie derivative of the metric is independent of the choice of connection,
\begin{equation}
\mathcal{L}_\xi g_{\mu\nu}
=
\xi^\alpha \partial_\alpha g_{\mu\nu}
+
g_{\alpha\mu}\partial_\nu\xi^\alpha
+
g_{\alpha\nu}\partial_\mu\xi^\alpha .
\end{equation}
To extract a conservation law we rewrite this expression using a covariant derivative.
If the Levi-Civita derivative is used,
\begin{equation}
\mathcal{L}_\xi g_{\mu\nu}
=
\mathring{\nabla}_\mu \xi_\nu
+
\mathring{\nabla}_\nu \xi_\mu,
\end{equation}
the variation becomes
\begin{equation}
\delta_\xi S_m
=
\int \d^4x\sqrt{-g}
T^{\mu\nu}
\mathring{\nabla}_\mu \xi_\nu .
\end{equation}
Integrating by parts gives
\begin{equation}
\delta_\xi S_m
=
\int \d^4x
\left[
\mathring{\nabla}_\mu
(\sqrt{-g}T^{\mu\nu}\xi_\nu)
-
\xi_\nu
\mathring{\nabla}_\mu(\sqrt{-g}T^{\mu\nu})
\right].
\end{equation}

For the Levi-Civita connection $\mathring{\nabla}_\mu g=0$, implying $\mathring{\nabla}_\mu\sqrt{-g}=0$. The first term therefore reduces to a boundary term that vanishes under standard assumptions. One obtains
\begin{equation}
\delta_\xi S_m
=
-\int \d^4x\sqrt{-g}\,
\xi_\nu
(\mathring{\nabla}_\mu T^{\mu\nu}).
\end{equation}
Requiring the variation to vanish for arbitrary $\xi^\mu$ yields
\begin{equation}
\mathring{\nabla}_\mu T^{\mu\nu}=0 .
\end{equation}

If instead the affine derivative $\nabla_\mu$ associated with $\tensor{\Gamma}{^\lambda_\mu_\nu}$ is used, the same Lie derivative should be rewritten with care. Since the independent connection is torsion-free in the symmetric teleparallel gravity, and since our convention is
$Q_{\alpha\mu\nu}=\nabla_\alpha g_{\mu\nu}$, we obtain
\begin{equation}
\mathcal{L}_\xi g_{\mu\nu}
=
Q_{\lambda\mu\nu}\xi^\lambda
+
2g_{\lambda(\nu}\nabla_{\mu)}\xi^\lambda .
\end{equation}
Therefore the variation of the matter action can be written as
\begin{equation}
\delta_\xi S_m
=
\int \d^4x\sqrt{-g}
\left(
T^\mu{}_\lambda\nabla_\mu\xi^\lambda
+
\frac{1}{2}
T^{\mu\nu}Q_{\lambda\mu\nu}\xi^\lambda
\right),
\end{equation}
where $T^\mu{}_\lambda\equiv T^{\mu\nu}g_{\nu\lambda}$ and the symmetry of $T^{\mu\nu}$ has been used. The first term can be integrated by parts. Using
\begin{equation}
\nabla_\mu\sqrt{-g}
=
\frac{1}{2}\sqrt{-g}Q_\mu,
\qquad
Q_\mu\equiv\tensor{Q}{_\mu_\lambda^\lambda},
\end{equation}
we derive
\begin{equation}
\int \d^4x\sqrt{-g}\,
T^\mu{}_\lambda\nabla_\mu\xi^\lambda
=
-\int \d^4x\sqrt{-g}\,
\xi^\lambda
\left(
\nabla_\mu T^\mu{}_\lambda
+
\frac{1}{2}Q_\mu T^\mu{}_\lambda
\right),
\end{equation}
up to a boundary term. Combining the two contributions gives
\begin{equation}
\delta_\xi S_m
=
\int \d^4x\sqrt{-g}\,
\xi^\lambda
\left[
-\nabla_\mu T^\mu{}_\lambda
-
\frac{1}{2}Q_\mu T^\mu{}_\lambda
+
\frac{1}{2}T^{\mu\nu}Q_{\lambda\mu\nu}
\right].
\end{equation}
Requiring $\delta_\xi S_m=0$ for arbitrary $\xi^\lambda$ then leads to the affine-connection form of the diffeomorphism identity,
\begin{equation}
\nabla_\mu T^\mu{}_\lambda
=
\frac{1}{2}T^{\mu\nu}Q_{\lambda\mu\nu}
-
\frac{1}{2}Q_\mu T^\mu{}_\lambda .
\end{equation}

This equation should not be interpreted as an independent force law. It is the same diffeomorphism identity expressed in terms of the non-metric affine derivative. The non-metricity terms appear because $\nabla_\alpha g_{\mu\nu}\neq0$. For minimally coupled matter, the physical conservation law is therefore most transparently written in the metric-compatible form
$\mathring{\nabla}_\mu T^{\mu\nu}=0$. For freely falling test particles and for pressureless dusts, this implies Levi-Civita geodesic motion after the continuity equation is used.

Finally, it is worth emphasizing that the geodesic criterion employed in this work depends only on the metric and is therefore independent of the choice of connection gauge. In particular, it remains valid in the coincident gauge used throughout this work.

\section{Proof that the second junction condition reproduces Israel’s second condition}

Consider the metric field equation in $f(\mathbb{Q})$ gravity
\begin{equation}
\frac{2}{\sqrt{-g}}\nabla_\alpha
\left[\sqrt{-g}f'(\mathbb{Q})\tensor{P}{^\alpha_\mu_\nu}\right]
+f'(\mathbb{Q})q_{\mu\nu}
-\frac{1}{2}f(\mathbb{Q})g_{\mu\nu}
=
\kappa T_{\mu\nu}.
\end{equation}
To analyze the junction conditions, we perform an infinitesimal integration of the field equations across the hypersurface $\Sigma$. We assume that the induced metric is continuous across $\Sigma$, while $Q_{\alpha\mu\nu}$ and $\tensor{P}{^\alpha_\mu_\nu}$ are piecewise regular and may have finite jumps. Then only the divergence term produces a distributional contribution, yeilding

\begin{equation}
\llbracket
n_\alpha f'(\mathbb{Q})\tensor{P}{^\alpha_\mu_\nu}
\rrbracket
=
\frac{\kappa}{2}S_{\mu\nu},
\end{equation}
where $S_{\mu\nu}$ denotes the surface stress-energy tensor on the boundary and $\llbracket\cdot\rrbracket$ represents the discontinuity across the hypersurface,
\begin{equation}
\llbracket A\rrbracket \equiv A|_{\Sigma^+}-A|_{\Sigma^-},
\end{equation}
with $\Sigma^\pm$ denoting the two sides of $\Sigma$.

To make contact with the standard junction conditions in GR, we introduce Gaussian normal coordinates near the hypersurface,

\begin{equation}
\d s^2=\d t^2+h_{ij}(t,x^k)\d x^i \d x^j,
\end{equation}
with unit normal vector
\begin{equation}
n^\mu=(1,0,0,0)^\top,\qquad n_\mu=(1,0,0,0).
\end{equation}
In GR, the extrinsic curvature of the hypersurface is defined as
\begin{equation}
K_{ij}=\frac{1}{2}\partial_t g_{ij}.
\end{equation}
We further introduce the traces of the non-metricity tensor,
\begin{equation}
Q_\alpha=\tensor{Q}{_\alpha_\lambda^\lambda},\qquad
\overline{Q}_\alpha=\tensor{Q}{^\lambda_\alpha_\lambda}.
\end{equation}
Working in the coincident gauge, we reduce the tensor $\tensor{P}{^\alpha_\mu_\nu}$  to
\begin{equation}
\tensor{P}{^t_i_j}
=
-\frac{1}{4}\tensor{Q}{^t_i_j}
+\frac{1}{2}\tensor{Q}{_(_i^t_j_)}
+\frac{1}{4}Q^{t}g_{ij}
-\frac{1}{4}\overline{Q}^{t}g_{ij}
-\frac{1}{4}\tensor{\delta}{^t_(_i}Q_{j)}.
\end{equation}
Using the relation between non-metricity and the time derivative of the spatial metric, we simplify the above expression,
\begin{equation}
\tensor{P}{^t_i_j}
=
-\frac{1}{2}K_{ij}
+
\frac{1}{2}K h_{ij}
=
-\frac{1}{2}(K_{ij}-K h_{ij}),
\end{equation}
where $K=h^{ij}K_{ij}$.

Now we consider the special case $f(\mathbb{Q})=\mathbb{Q}$, for which $f'(\mathbb{Q})=1$. In the absence of surface matter ($S_{\mu\nu}=0$), the junction condition becomes
\begin{equation}
\llbracket
n_\alpha f'(\mathbb{Q})\tensor{P}{^\alpha_\mu_\nu}
\rrbracket
=0
\quad\Longrightarrow\quad
\llbracket\tensor{P}{^t_i_j}\rrbracket=0.
\end{equation}
Substituting the above expression for $\tensor{P}{^t_i_j}$ gives
\begin{equation}
\llbracket K_{ij}-K h_{ij}\rrbracket=0,
\end{equation}
which is precisely Israel’s second junction condition.
For convenience we introduce
\begin{equation}
\Pi^{\alpha\mu\nu}
=
\sqrt{-g}\,f'(\mathbb{Q})P^{\alpha\mu\nu}.
\end{equation}
With this definition, our second junction condition can be written as
\begin{equation}
\llbracket n_\alpha\tensor{\Pi}{^\alpha_\mu_\nu}\rrbracket=0.
\end{equation}
Since $\Pi^{\alpha\mu\nu}$ satisfies
\begin{equation}
\Pi^{\alpha\mu\nu}
=
\frac{\partial \mathcal{L}}{\partial Q_{\alpha\mu\nu}},
\qquad
\mathcal{L}=\frac{\sqrt{-g}}{2\kappa}f(\mathbb{Q}),
\end{equation}
this condition can be interpreted as the conservation of the canonical momentum conjugate to the metric degrees of freedom.

\section{Proof that the affine parameter integral for null geodesics is finite in the saturated source of Scenario 1}

We prove that the affine parameter for null geodesics remains finite when the source saturates in Scenario~1.

Assume that
\begin{equation}
\lim_{t\to\infty}\frac{\dot{a}(t)}{a(t)}=-H_0,
\qquad
H_0=\sqrt{\frac{\mathbb{Q}_0}{6}}>0 .
\end{equation}
The affine parameter along a radial null geodesic satisfies
\begin{equation}
\lambda \propto \int_{t_0}^{\infty} a(t)\,\mathrm{d}t .
\end{equation}
We show that this integral converges.

\vspace{0.3cm}

From the definition of the limit, for any $\epsilon>0$ there exists a time $T>t_0$ such that for all $t>T$,
\begin{equation}
\left|
\frac{\dot{a}(t)}{a(t)}+H_0
\right|
<\epsilon ,
\end{equation}
which implies
\begin{equation}
-H_0-\epsilon
<
\frac{\dot{a}(t)}{a(t)}
<
-H_0+\epsilon .
\end{equation}
Choosing $\epsilon=H_0/2$, we obtain
\begin{equation}
\frac{\dot{a}(t)}{a(t)}<-\frac{H_0}{2},
\qquad t>T .
\end{equation}
Integrating this inequality from $T$ to $t$ gives
\begin{equation}
\int_T^t
\frac{\mathrm{d}}{\mathrm{d}\tau}
\left(\ln a(\tau)\right)\mathrm{d}\tau
<
\int_T^t
-\frac{H_0}{2}\,\mathrm{d}\tau ,
\end{equation}
which yields
\begin{equation}
\ln a(t)-\ln a(T)
<
-\frac{H_0}{2}(t-T).
\end{equation}
Exponentiating both sides gives the bound
\begin{equation}
a(t)
<
a(T)\exp\!\left(\frac{H_0}{2}T\right)
\exp\!\left(-\frac{H_0}{2}t\right).
\end{equation}
Defining
\begin{equation}
C=a(T)\exp\!\left(\frac{H_0}{2}T\right),
\end{equation}
we obtain
\begin{equation}
0<a(t)<C\exp\!\left(-\frac{H_0}{2}t\right),
\qquad \forall\,t>T,
\end{equation}
where $0<C<\infty$.
The affine parameter can now be written as
\begin{equation}
\lambda \propto
\int_{t_0}^{\infty} a(t)\,\mathrm{d}t
=
\int_{t_0}^{T} a(t)\,\mathrm{d}t
+
\int_{T}^{\infty} a(t)\,\mathrm{d}t .
\end{equation}
The first term is finite because $a(t)$ is finite on the compact interval $[t_0,T]$,
\begin{equation}
\int_{t_0}^{T} a(t)\,\mathrm{d}t < \infty .
\end{equation}
For the second term, using the exponential bound above,
\begin{equation}
\int_{T}^{\infty} a(t)\,\mathrm{d}t
<
\int_{T}^{\infty}
C\exp\!\left(-\frac{H_0}{2}t\right)\mathrm{d}t
=
\frac{2C}{H_0}
\exp\!\left(-\frac{H_0}{2}T\right)
<\infty,
\end{equation}
we therefore obtain
\begin{equation}
\lambda \propto \int_{t_0}^{\infty} a(t)\,\mathrm{d}t < \infty ,
\end{equation}
which completes the proof.

\section{Proof that bounce solutions satisfy the regularity criteria}

Let $a(t)$ be a bounce solution of the modified Friedmann dynamics. 
A bounce configuration implies that the scale factor admits a strictly positive lower bound,
\begin{equation}
    a(t) \ge a_{\min} > 0, \qquad \forall\, t\in(-\infty,\infty).
\end{equation}
We now verify that such solutions satisfy the two regularity criteria introduced in the main text.

\subsubsection{Geometric Finiteness}

In the coincident gauge of the FLRW metric, the non-metricity scalar reads
\begin{equation}
\mathbb{Q}=6H^2=6\left(\frac{\dot a}{a}\right)^2 .
\end{equation}
Since $a(t)$ never approaches zero and the evolution is assumed to be smooth, the derivative $\dot a(t)$ remains finite for solutions generated by the master equation
\[
\mathcal{J}(\mathbb{Q})=\epsilon_{\text{eff}}(a),
\]
where $\epsilon_{\text{eff}}(a)$ is finite for $a\ge a_{\min}$. Consequently, the scalar $\mathbb{Q}$ is finite for all cosmic times.

Higher-order invariants constructed from $\mathbb{Q}$ and its covariant derivatives involve terms of the schematic form $\ddot a/a$ and $(\dot a/a)^2$. The condition $a(t)\neq0$ guarantees that these denominators never vanish. Together with the smoothness of the solution, this ensures that all algebraic and differential invariants constructed from $\mathbb{Q}$ remain finite. Hence the spacetime satisfies the geometric finiteness criterion.

\subsubsection{Geodesic Completeness}

Consider the affine parameter $\lambda$ along a radial null geodesic. In FLRW spacetime the relation between the affine parameter and the coordinate time satisfies $\mathrm{d}\lambda \propto a(t)\mathrm{d}t$. Integrating from an initial time $t_0$ to $t$ yields
\begin{equation}
\lambda(t)\propto \int_{t_0}^{t} a(\tau)\,\mathrm{d}\tau .
\end{equation}
Using the lower bound $a(t)\ge a_{\min}$ we obtain
\begin{equation}
\int_{t_0}^{t} a(\tau)\,\mathrm{d}\tau
\ge
\int_{t_0}^{t} a_{\min}\,\mathrm{d}\tau
=
a_{\min}(t-t_0).
\end{equation}
Taking the limit $t\to\infty$ and noting that $a_{\min}>0$, we find $\lambda(t)\to\infty$. Therefore, the null geodesics can be extended to arbitrarily large affine parameter. An analogous argument applies to timelike geodesics. Hence all physical observers can extend their worldlines indefinitely, implying geodesic completeness.

\vspace{0.3cm}

We therefore conclude that the existence of a strictly positive lower bound $a_{\min}$ is sufficient to ensure both intrinsic regularity criteria. In particular, any bounce solution generated by the non-perturbative dynamics considered here corresponds to a physically regular spacetime.

\section{Derivation of the asymptotic solution in the quadratic analytic model}
We derive the asymptotic expression used in Eq.~\ref{eq:approx_a}.
For $a\to 0$,
\begin{equation}
    \ln\left(\frac{3m_0\xi}{a^3}+1\right)\simeq\ln\left(\frac{3m_0\xi}{a^3}\right).
\end{equation}
Equation~\ref{eq:quad_adott} then gives
\begin{equation}
    \frac{\dot a}{a}\simeq-\frac{1}{6}\left[\frac{24}{\alpha\xi}\ln\left(\frac{3m_0\xi}{a^3}\right)\right]^{1/4}.
\end{equation}

Define
\begin{equation}
    D_0=(3m_0\xi)^{1/3},\qquad x=\ln\frac{D_0}{a},
\end{equation}
then
\begin{equation}
    \ln\left(\frac{3m_0\xi}{a^3}\right)=3x,\qquad \dot x=-\frac{\dot a}{a}.
\end{equation}
Therefore
\begin{equation}
    \dot x\simeq\frac{1}{6}\left(\frac{24}{\alpha\xi}\right)^{1/4}3^{1/4}x^{1/4}.
\end{equation}

Integration gives
\begin{equation}
    x^{3/4}\simeq\frac{3}{4}\cdot\frac{1}{6}\left(\frac{24}{\alpha\xi}\right)^{1/4}\cdot3^{1/4}(t-t_1),
\end{equation}
or, equivalently,
\begin{equation}
    x\simeq E_0(C_0t-C_1)^{4/3},
\end{equation}
where
\begin{equation}
    C_0=\frac{1}{6}\left(\frac{24}{\alpha\xi}\right)^{1/4},\qquad E_0=\left(\frac{3^{5/4}}{4}\right)^{4/3}=\left(\frac{3^5}{2^8}\right)^{1/3}.
\end{equation}

Since $a=D_0e^{-x}$, we obtain
\begin{equation}
    a(t)\sim D_0\exp\left[-E_0(C_0t-C_1)^{4/3}\right],
\end{equation}
which is Eq.~\ref{eq:approx_a}.

\section{Derivation of the exterior mass function}

We derive Eq.~\ref{eq:exterior_mass_func} from the intrinsic junction
conditions. On the boundary of the dust cloud, the areal radius is
\begin{equation}
    R(t)=r_b a(t),
\end{equation}
where $r_b$ is the comoving radius of the boundary. 

For the exterior region we write the static spherically symmetric line element in areal-radius coordinates, so that the metric function can be
parametrized by an exterior mass function $M(R)$ through $F(R)=1-2M(R)/R$. On the boundary $\Sigma$ of the dust cloud, the first junction condition identifies the areal radius of the exterior geometry with that of the FLRW interior,
\begin{equation}
    R_\Sigma(t)=r_b a(t).
\end{equation}

The second junction condition then fixes the normal momentum flux across $\Sigma$. By evaluating the second junction condition for the FLRW interior and assuming that $\mathbb{Q}$ is continuous across the boundary with $f'(\mathbb{Q}_\Sigma)\neq 0$, considering outside metric function as $F(R)=1-2M(R)/R$, we find
\begin{equation}
    \frac{2M(R_\Sigma)}{R_\Sigma}=\dot R_\Sigma^{\,2}.
\end{equation}
Since $\dot R_\Sigma=r_b\dot a=R_\Sigma H$, we obtain
\begin{equation}
    M(R_\Sigma)=\frac{R_\Sigma^3H^2}{2}.
\end{equation}

After the matching is imposed, we drop the subscript $\Sigma$ and regard
this as the exterior mass function evaluated at the boundary radius,
\begin{equation}
    M(R)=\frac{R^3H^2}{2}.
\end{equation}

From Eq.~\ref{eq:quad_adott},
\begin{equation}
    H^2=\left(\frac{\dot a}{a}\right)^2=\frac{1}{36\alpha}\left[\sqrt{\frac{24\alpha}{\xi}\ln\left(\frac{3m_0\xi}{a^3}+1\right)+1}-1\right],
\end{equation}
and
using $a=R/r_b$, we derive
\begin{equation}
    H^2(R)=\frac{1}{36\alpha}\left[\sqrt{\frac{24\alpha}{\xi}\ln\left(\frac{3m_0\xi r_b^3}{R^3}+1\right)+1}-1\right].
\end{equation}
Substituting the above equation into $M(R)=R^3H^2/2$ gives
\begin{equation}
    M(R)=\frac{R^3}{72\alpha}\left[\sqrt{\frac{24\alpha}{\xi}\ln\left(\frac{3m_0\xi r_b^3}{R^3}+1\right)+1}-1\right],
\end{equation}
which is Eq.~\ref{eq:exterior_mass_func}.

\end{document}